\documentclass[showpacs,floats,floatfix,aps,prb,groupedaddress,twocolumn,amsfonts,amssymb]{revtex4}

\usepackage{bm}
\usepackage{hyphenat}
\usepackage{graphicx}
\usepackage{psfrag}
\usepackage{psfig}
\usepackage[latin1]{inputenc}
\begin{document}
\renewcommand{\thefootnote}{\fnsymbol{footnote}}

\title{Vortex-line percolation in the three-dimensional\\ complex $|\psi|^4$ model}
\author{Elmar Bittner, Axel Krinner and Wolfhard Janke}

\affiliation{Institut f\"ur Theoretische Physik, Universit\"at Leipzig,
Augustusplatz 10/11, D-04109 Leipzig, Germany}

\begin{abstract}
\noindent 
In discussing the phase transition of the three-dimensional complex $|\psi|^4$ 
theory, we
study the geometrically defined vortex-loop network as well as the
magnetic properties of the system in the vicinity of the critical point.
Using high-precision Monte Carlo techniques 
we investigate if both of them exhibit the same
critical behavior leading to the same critical exponents and hence to
a consistent description of the phase transition.
Different percolation observables are taken into account
and compared with each other. We find that different connectivity definitions 
for constructing the vortex-loop network lead to different results in the 
thermodynamic limit, and the percolation thresholds do not coincide with the 
thermodynamic phase transition point.
\end{abstract}


\maketitle

\section{Introduction} \label{intro} 

Substantial progress in the understanding of the nature of phase transitions
driven by topological excitations has been achieved in the beginning of the 
1970's, when Berezinskii~\cite{berenzinskii} and
Kosterlitz and Thouless~\cite{KosterlitzThouless} published their seminal papers
on the two-dimensional $XY$ model, involving the unbinding of point like vortices 
when the temperature exceeds a critical value. A few years later in 1977, 
Banks, Myerson and Kogut~\cite{BanksKogutMyerson} showed
that the Villain model in three dimensions, a particular
spin model with global $O(2)$ symmetry due to the $2\pi$-periodicity in the
Hamiltonian, can be represented by an equivalent defect model with
long-range Biot-Savart-like interactions, where the spin configurations are 
integer valued and sourceless. These configurations can be interpreted as 
line like excitations forming closed networks which can be identified with the
vortex loops of the original theory. At the transition point, where the
broken $O(2)$ symmetry in the low-temperature phase is restored, loops of 
infinite length become important which provides the basis for attempting a 
percolational treatment.\cite{stauffer} So the question arises whether the 
percolational threshold coincides with the thermodynamic critical point, or 
under which conditions such a coincidence can be established.~\cite{kajantie}

Percolational studies of spin clusters in the
Ising model showed that one has to handle this approach carefully. It
only works, if one uses a proper stochastic definition of clusters.~\cite{kasteleyn,coniglio,Fortunato,wjas}
Such so-called Fortuin-Kasteleyn clusters of spins can be obtained from the geometrical spin
clusters, which consist of nearest-neighbor sites with their spin variables in the same
state, by laying bonds with a certain probability between the nearest neighbors. 
The resulting Fortuin-Kasteleyn clusters are in general smaller than the 
geometrical ones and also more
loosely connected. It is well known that depending on the cluster type considered
one may find different sets of critical exponents and even different percolation 
thresholds, so a careful treatment is required.

In three-dimensional, globally $O(2)$ symmetric theories the percolating
objects are vortex lines forming closed networks.
The question we want to address in this paper is:
Is there a similar clue in the case of vortex networks as for spin clusters,
or do they display different features?
Therefore we connect the obtained vortex line elements to closed loops, which are 
geometrically defined objects. When a branching point, where $n\ge2$ junctions are 
encountered, is reached, a decision on how to continue has to be made. This step 
involves a certain ambiguity.
We want to investigate the influence of the probability of treating such a 
branching point
as a knot. This work concentrates on the three-dimensional complex 
Ginzburg-Landau model, the field theoretical representative of the $O(2)$ 
universality class. 

The rest of the paper is organized as follows. In Sec.~\ref{model} we give the 
definition of the model and introduce the observables. The results of our 
Monte Carlo simulations are presented in Sec.~\ref{results}, and concluding 
remarks and an outlook to future work can be found in Sec.~\ref{conclusion}.

\vspace{-1mm}
\section{Model and Observables} \label{model} 

The standard complex or two-component Ginzburg-Landau theory is defined by the
Hamiltonian
\begin{equation}
H[\psi] = \int \!\! \mathrm{d}^dr \left[\alpha |\psi|^2 + \frac{b}{2}|\psi|^4 +
\frac{\gamma}{2}|\nabla \psi|^2 \right], \quad \gamma > 0~,
\label{eq:H}
\end{equation}
where $\psi(\vec{r}) = \psi_x(\vec{r}) + i  \psi_y(\vec{r}) = 
|\psi(\vec{r})| e^{i \phi(\vec{r})}$ is a complex
field, and $\alpha$, $b$ and $\gamma$ are temperature independent coefficients 
derived from a microscopic model.
In order to carry out Monte Carlo simulations we put
the model (\ref{eq:H}) on a $d$-dimensional hypercubic lattice with spacing $a$.
Adopting the notation of Ref.~\onlinecite{beck1}, we introduce scaled variables
$\tilde{\psi} = \psi/\sqrt{(|\alpha|/b)}$ and $\vec{u}=\vec{r}/ \xi$,
where $\xi^2=\gamma/|\alpha|$ is the mean-field correlation length at zero
temperature. This leads to the normalized lattice Hamiltonian
\begin{equation}\label{h2}
H[\tilde{\psi}] = k_B \tilde{V}_0  \sum_{n=1}^N \Big [\frac{\tilde{\sigma}}{2}
(|\tilde{\psi}_n|^2 - 1)^2 +
\frac{1}{2}\sum_{\mu=1}^d |\tilde{\psi}_n-\tilde{\psi}_{n+\mu}|^2 \Big ]~,
\end{equation}
with 
\begin{equation}
\tilde{V}_0=\frac{1}{k_B}\frac{|\alpha|}{b}\gamma a^{d-2}~,\quad
\tilde{\sigma}=\frac{a^2}{\xi^2}~,
\end{equation}
where $\mu$ denotes the unit vectors along the $d$ coordinate axes, 
$N=L^d$ is the total
number of sites, and an unimportant constant term has been dropped. The
parameter $\tilde{V}_0$ merely sets the temperature scale and can thus be
absorbed in the definition of the reduced temperature 
$\tilde{T} = T/\tilde{V}_0$.

After these rescalings and omitting the tilde on $\psi$, $\sigma$, and $T$ 
for notational simplicity in the rest of the paper,
the partition function $Z$ considered in the simulations is given by
\begin{equation}
Z=\int \!\! D\psi D\bar{\psi} \, e^{-\beta H}~,
\label{eq:Z}
\end{equation}
where $\beta =1/T$ denotes the inverse temperature and
$\int D\psi \,D\bar{\psi} \equiv \int D\,{\rm Re\/}\psi \,D\,{\rm Im\/}\psi$
stands short for integrating over all possible complex field configurations.

In the limit of a large parameter $\sigma$, it is easy to read off from 
Eq.~(\ref{h2}) that the modulus of the field is squeezed onto unity such
that  the $XY$ model limit is approached with its well-known continuous 
phase transition in three dimensions at $\beta_c \approx 0.45$.~\cite{WJ_XY}

In order to characterize the transition we have measured 
in our simulations to be described in detail in the next section
among other quantities the energy $\langle H\rangle$,
the specific heat $c_v=(\langle H^2\rangle- \langle H\rangle^2)/N$, and
the mean-square amplitude $\langle|\psi|^2\rangle = (1/N) \sum_{n=1}^N
\langle |\psi_n|^2\rangle$.
In order to determine the critical temperature, the helicity
modulus,
\begin{eqnarray}
\Gamma_\mu = \frac{1}{N}\langle \sum_{n=1}^N
|\psi_n||\psi_{n+\mu}|
\cos(\phi_n - \phi_{n+\mu})\rangle \nonumber\\
-\frac{1}{NT} \langle \left[\sum_{n=1}^N|\psi_n|
|\psi_{n+\mu}| \sin(\phi_n - \phi_{n+\mu}) \right]^2 \rangle~,
\label{eq:helicity}
\end{eqnarray}
and the Binder cumulant 
$U = \langle |M|^4 \rangle/\langle |M|^2 \rangle^2$
were also computed, where $M = M_x + i M_y = \sum_{n=1}^N \psi_n$
is the magnetization of a given configuration.

The main focus in this  paper is on the properties of the geometrically 
defined vortex-loop network.
The standard procedure to calculate
the vorticity on each plaquette is by considering the quantity 
\begin{equation}
m=\frac{1}{2\pi}([\phi_1-\phi_2]_{2\pi}+[\phi_2-\phi_3]_{2\pi}+[\phi_3-\phi_4]_{2\pi}+[\phi_4-\phi_1]_{2\pi})~,
\end{equation}
where $\phi_1,\dots,\phi_4$ are the phases at the corners of a plaquette
labeled, say, according to the right-hand rule, and $[\alpha]_{2\pi}$ 
stands for $\alpha$ modulo $2\pi$: $[\alpha]_{2\pi}=\alpha+2\pi n$,
with $n$ an integer such that $\alpha+2\pi n \in (-\pi,\pi]$, hence 
$m=n_{12}+n_{23}+n_{34}+n_{41}$. If $m\neq0$, there exists a topological
charge which is assigned to the object dual to the given plaquette, i.e.,
the (oriented) line elements ${*l_\mu}$ which combine to form closed networks 
(``vortex loops''). With this definition, the vortex ``currents'' ${*l_\mu}$
can take three values: $0,\pm 1$ (the values $\pm 2$ have a 
negligible probability and higher values are impossible). The quantity
\begin{eqnarray} 
v &=& \frac{1}{N}\sum_{n,\mu}|{*l}_{\mu,n}| 
\label{eq:vortex}
\end{eqnarray}
serves as a measure of the vortex-line density.

\begin{figure}[t]
\centerline{\psfig{figure=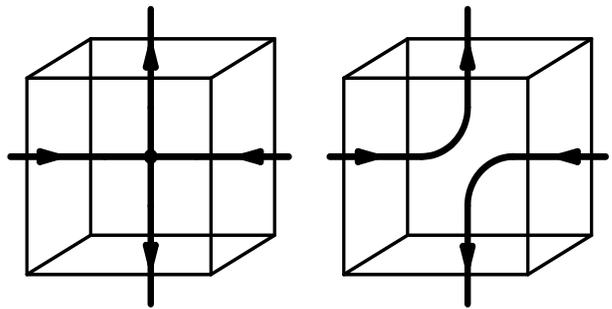,angle=0,height=4.cm,width=8.cm}}
\caption{\label{VortexCon}
If two (or three) vortex lines pass through one cell, the vortex tracing 
algorithm must decide how to connect them, and this leads to an ambiguity 
in the length distribution. Left: Connecting all line elements (forming a knot).
Right: Connections are made stochastically.
}
\end{figure}

In order to study percolation observables we connect the
obtained vortex line elements to closed loops, which are geometrically
defined objects. Following a single line, there is evidently no
difficulty, but when a branching point, where $n\ge2$ junctions are encountered,
is reached, a decision on how to continue has to be made.
This step involves a certain ambiguity. If we connect all in- and out-going
line elements, knots will be formed. Another choice is to join only 
one incoming with one outgoing line element, with the outgoing
direction chosen randomly. These two possibilities are shown in Fig.~\ref{VortexCon}.
We will employ two ``connectivity'' definitions here:

\begin{itemize}

\item ``Maximal'' rule: At all branching points, we 
  connect all line elements, such that the maximal loop length is
  achieved. That means each branching point is treated as a knot.

\item ``Stochastic'' rule: At a branching point where $n\ge2$
  junctions are encountered, we draw a uniformly distributed random number
  $\in (0,1]$ and if this
  number is smaller than the {\it connectivity} parameter $c$ we 
  identify this branching point as a knot of the loop,
  i.e., only with probability $0 \le c \le 1$ a branching point is treated 
  as a knot. In this way we can systematically interpolate between the
  maximal rule for $c=1$ and the case $c=0$, which corresponds 
  to the procedure most commonly followed in the literature.~\cite{kajantie}

\end{itemize}

\begin{figure*}
\centerline{\psfig{figure=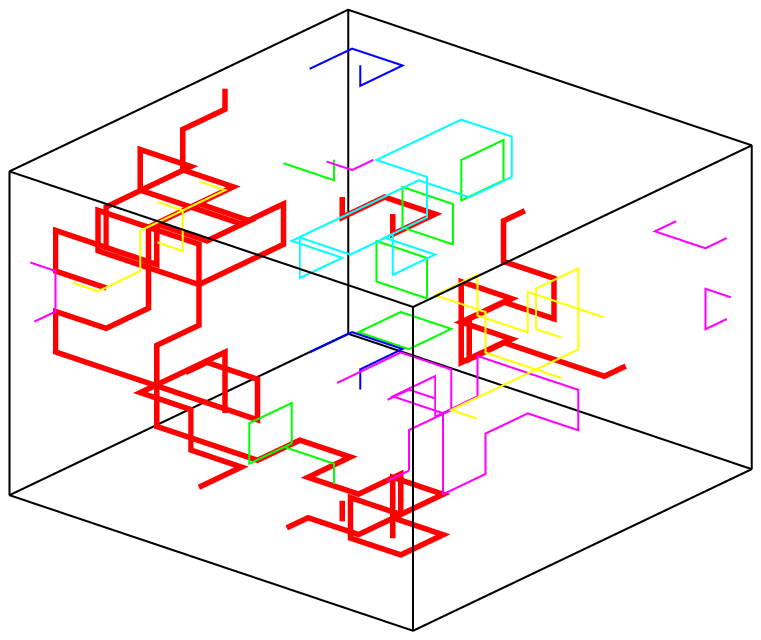,angle=0,height=6.5cm,width=7cm        }\hspace{1cm}
\psfig{figure=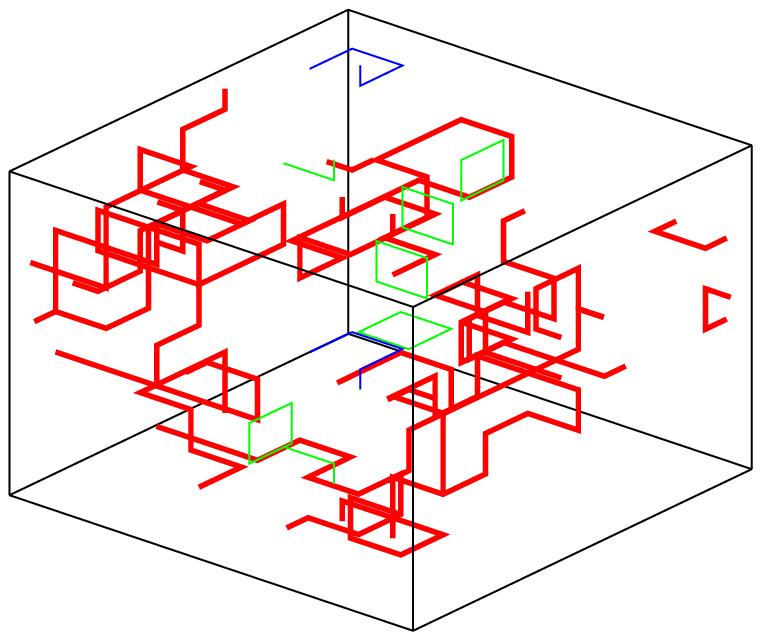,angle=0,height=6.5cm,width=7cm}}
\caption{
\label{fig:3d}
(Color online) Vortex-loop networks for the ``stochastic'' definition 
with $c=0.4$ (left) and the ``maximal'' definition with $c=1$ (right). Both networks are 
generated from the same $L=8$ lattice field configuration at the (thermodynamically) 
critical coupling $\beta_c=0.780\,08$. The different loops are distinguished by the 
color coding.
}
\end{figure*}

We can thus extract from each lattice configuration a set of vortex loops,
which have been glued together by one of the connectivity definitions above. 
In Fig.~\ref{fig:3d} we show two possible vortex-loop networks for $c=0.4$ and
$c=1$ generated out of the same lattice configuration.

For each loop in the network, we measure the following observables:
 
\begin{itemize}

\item ``Mass'', ${\cal O}_{\rm mass}$:  
The ``mass'' of a vortex loop is the number of line elements $*l_{\mu,n}$
 of the loop, i.e., simply its length $l_{\rm loop}$ normalized by the volume
\begin{equation}
\mathcal{O}_{\rm{mass}} \equiv l_{\rm loop}/N~.
\end{equation}
By summing over all loops of a configuration we recover of course the vortex
density~(\ref{eq:vortex}),
\begin{equation}
\sum_{\rm loops}{O}_{\rm{mass}} = v~.  
\end{equation}
For the percolation analysis the mass of the longest loop 
$\mathcal{O}_{\rm{mass}}^{\rm max}$ in each vortex network is recorded, which
 usually serves as a  measure of the percolation strength (behaving similarly 
to a magnetization).~\cite{stauffer}

\item ``Volume'', ${\cal O}_{\rm vol}$: 
For each vortex loop, first the smallest rectangular box is determined that 
contains the whole loop. This value is then normalized by the volume of the
lattice. A vortex loop spread over an extent $l_x$, $l_y$, and $l_z$ thus 
results in
\begin{equation}
\mathcal{O}_{\rm{vol}}=(l_x\times l_y\times l_z)/N~.
\end{equation}
For each lattice configuration, we record the maximal ``volume'' 
$\mathcal{O}_{\rm{vol}}^{\rm max}$, which may be taken 
as an alternative definition of the percolation strength.

\item ``Extent'' of a vortex loop in 1, 2, or 3 dimensions, ${\cal O}_{\rm 1D}, 
{\cal O}_{\rm 2D},$ and ${\cal O}_{\rm 3D}$:
This means simply to project the loop onto the three axes and record whether
the projection covers the whole axis, or to be more concrete, whether one
finds a vortex-line element of the loop in all planes perpendicular to the eyed
axis. If there is a loop fulfilling this requirement, then this loop is percolating
and we record $1$ in
the time series of measurements; if not, a value of $0$ is stored. This
quantity can thus be interpreted as percolation probability~\cite{stauffer} which
(behaving similarly to a Binder parameter) is a convenient quantity for
locating the percolation threshold $\beta_p$.

\item ``Susceptibilities,'' $\chi_i$:
For the vortex-line density $v$ and any of the observables ${\cal O}_i$ defined 
above ($i =$ ``mass,'' ``vol,'' ``1D,'' \dots), one can use its variance to define 
the associated susceptibility,
\begin{equation}
\chi_i=N (\langle {\cal O}_i^2 \rangle-\langle {\cal O}_i \rangle^2)~,
\label{eq:variances}
\end{equation}
which is expected to signal critical fluctuations.

\item ``Line tension,'' $\theta$: 
On general grounds the loop-length distribution $P(l_{\rm loop})$ is expected
 to have the following form:~\cite{adriaan}
\begin{equation}
P(l_{\rm loop})\sim l_{\rm loop}^{-\tau}\exp{(-l_{\rm loop}\theta)}~,
\label{eq:P_loop}
\end{equation}
where the Fisher exponent $\tau$ is given in terms of the fractal dimension 
$D$ of the loops by
\begin{equation}
\tau=\frac{d}{D}+1~.
\end{equation}
For a three-dimensional ($d=3$) (noninteracting) Brownian random walk with 
$D=2$ this leads to $\tau=5/2$, while $\tau > 5/2$ for self-avoiding and 
$\tau < 5/2$ for self-seeking lines, respectively. The parameter $\theta$ 
is the line tension which vanishes according to~\cite{kajantie}
\begin{equation}
\theta= |\beta-\beta_p|^{\gamma_{\theta}}~, 
\end{equation}
where $\beta_p$ is the percolation threshold of the random walk and 
$\gamma_{\theta} \equiv 1/\sigma_{\theta}$
the second independent percolation exponent.~\cite{stauffer}

\end{itemize}

\section{Simulation and Results} \label{results}

Let us now turn to the description of the Monte Carlo update procedures used
by us.  We employed the single-cluster algorithm~\cite{wolff} to update the 
direction of the field,~\cite{hasen} similar to simulations of the $XY$ spin 
model.~\cite{WJ_XY} The modulus of $\psi$ is updated with a Metropolis 
algorithm.~\cite{Metro,WJ_review} Here some care is necessary
to treat the measure in Eq.~(\ref{eq:Z}) properly (see Ref.~\onlinecite{ebwj_prl}).
One sweep consisted of $N$ spin flips with the Metropolis algorithm and $N_{\rm sc}$
single-cluster updates. For all simulations 
the number of cluster updates was chosen roughly proportional to the linear
lattice size, $N_{\rm sc} \simeq L$, a standard choice for three-dimensional 
systems as suggested by a simple finite-size scaling (FSS) argument.
We performed simulations for lattices with linear lattice size
$L=6$ -- $10,12,14,16,18,20,22,24,26,28,32,36,$ and  $40$, respectively,
subject to periodic boundary conditions.
After an initial equilibration time of $20\,000$ sweeps we took about
$100\,000$ measurements, with ten sweeps between the measurements.
All error bars are computed with the Jackknife method.~\cite{Jack}

\begin{table*}[htb]
\caption{\label{expos}The critical exponents of the 3D $XY$ model universality class
as reported in Ref.~\onlinecite{hasen2} and the correction-to-scaling exponent $\omega$ 
of Ref.~\onlinecite{hasen}.}
  \begin{center}
   \begin{tabular}{ccccccc}\hline\hline
    \makebox[2.cm][c]{$\alpha$} &\makebox[2.cm][c]{$\beta$} &\makebox[2.cm][c]{$\gamma$} 
     &\makebox[2.cm][c]{$\delta$}&\makebox[2.cm][c]{$\eta$} & \makebox[2.cm][c]{$\nu$}&\makebox[2.cm][c]{$\omega$}\\ \hline
    $-0.0146(8)$&0.3485(2)&1.3177(5)&4.780(2)&0.0380(4)&0.67155(27)&0.79(2)\\ \hline \hline
   \end{tabular}
  \end{center}
\end{table*}
In order to be able to compare standard, thermodynamically obtained 
results (working directly with the original field variables) with the 
percolative treatment of the geometrically defined vortex-loop networks
considered here, we used the same value for the parameter $\sigma=1.5$ 
as in Ref.~\onlinecite{ebwj_PRB} for which we determined
by means of standard FSS analyses a critical coupling of
\begin{equation}
\beta_c = 0.780\,08(4)~.
\label{eq:thermo_bc}
\end{equation}
Focussing here on the vortex loops, we performed new
simulations at this thermodynamically determined critical value, 
$\beta = 0.780\,08$, as well as additional simulations at $\beta=0.79$, $0.80$,
and $0.81$. The latter $\beta$ values were necessary because of the 
spreading of the pseudocritical points of the vortex loop related quantities. 
As previously we recorded the time series of the energy $H$,
the magnetization $M$, the mean modulus $\overline{|\psi|}$, and
the mean-square amplitude
$|\psi|^2$, as well as
the helicity modulus $\Gamma_\mu$ and the vortex-line density $v$. In the present 
simulations, however, we saved in addition also the field configurations in each 
measurement. This enabled us to perform the time consuming analyses of the 
vortex-loop networks after finishing the simulations and thus to systematically 
vary the connectivity parameter $c$ of the knots.

The FSS ansatz for the pseudocritical inverse temperatures $\beta_i(L)$, 
defined as the points where the various $\chi_i$ obtain their maxima, is 
taken as usual as 
\begin{equation}\label{fss_beta}
\beta_i(L) = \beta_{i,c} + c_1 L^{-1/\nu}+c_2 L^{-1/\nu-\omega} + \dots ~,
\end{equation} 
where $\beta_{i,c}$ denotes the infinite-volume limit, and $\nu$ and 
$\omega$ are the correlation length and confluent correction critical 
exponents, respectively. Here we have deliberately retained the subscript $i$
on $\beta_{i,c}$.

Let us start with the susceptibility $\chi_v$ of the vortex-line density. Note that 
this quantity, while also being expressed entirely in terms of vortex elements,
plays a special role in that it is locally defined, i.e., does {\em not\/} 
require a decomposition into individual vortex loops (which, in fact, is the 
time-consuming part of the vortex-network analysis). Assuming the 
$XY$ model values for $\nu$ and $\omega$ compiled in Table~\ref{expos}, which 
are taken from Refs.~\onlinecite{hasen} and \onlinecite{hasen2}, and fitting 
only the coefficients $\beta_{i,c}$ and $c_i$, we arrive at the estimate
\begin{equation}
\beta_{v,c} = 0.7797(14)~
\label{eq:vortex_bc}
\end{equation}
with a goodness-of-fit parameter $Q=0.20$. This value is perfectly consistent
with the previously obtained ``thermodynamic'' result (\ref{eq:thermo_bc}),
derived from FSS of the magnetic susceptibility and various (logarithmic) 
derivatives of the magnetization. On the basis of this result, it would be
indeed tempting to 
conclude that the phase transition in the three-dimensional complex 
Ginzburg-Landau field theory can be explained in terms of vortex-line 
{\em proliferation\/}.~\cite{antunes1,antunes2} As pointed out above, however,
the vortex-line density $v$ does not depend on the connectivity of the 
vortex network and therefore does not probe its percolation properties.
In fact, $v$ behaves similar to the energy and the associated
susceptibility $\chi_v$ similar to the specific heat, so that the good
agreement between Eqs.~(\ref{eq:thermo_bc}) and (\ref{eq:vortex_bc}) is {\em a priori} to
be expected.

\begin{figure*}[htb]
\vspace*{3cm}
\centerline{\psfig{figure=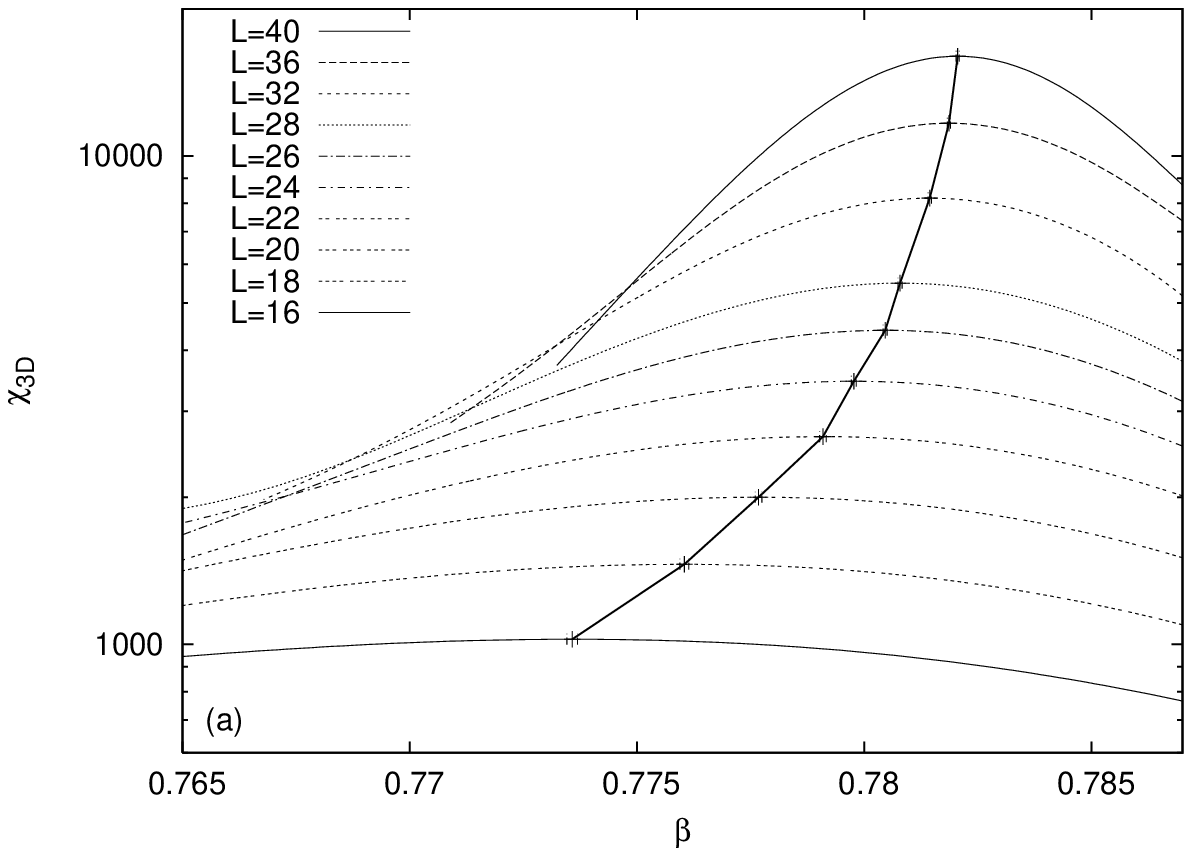,angle=0,height=5.5cm,width=7cm}
\psfig{figure=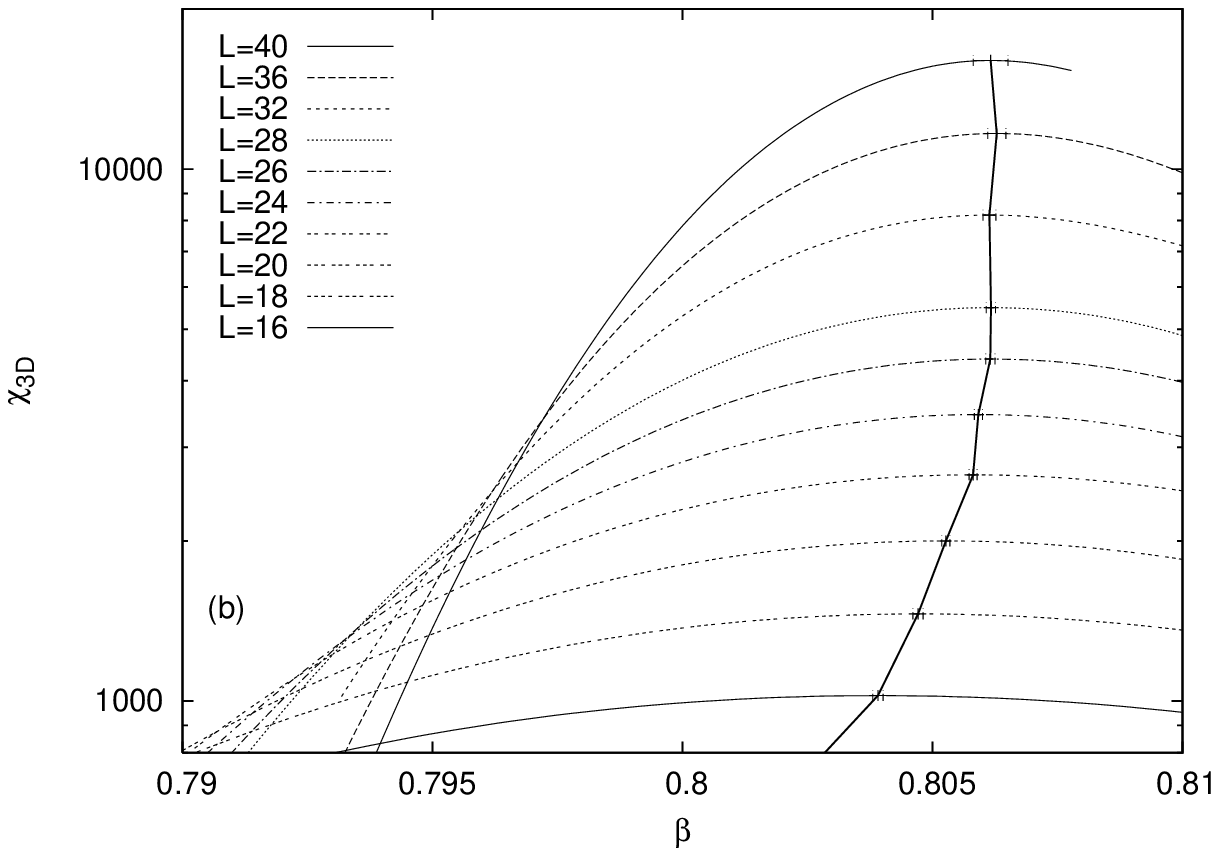,angle=0,height=5.5cm,width=7cm}}
\caption{
\label{fig:p3_0}
Susceptibility of ${\cal O}_{\rm 3D}$ as a function of inverse temperature $\beta = 1/T$ 
for (a) the ``stochastic'' rule ($c=0$) and (b) the ``maximal'' rule ($c=1$).
}
\end{figure*}

\begin{figure*}[htb]
\centerline{ \psfig{figure=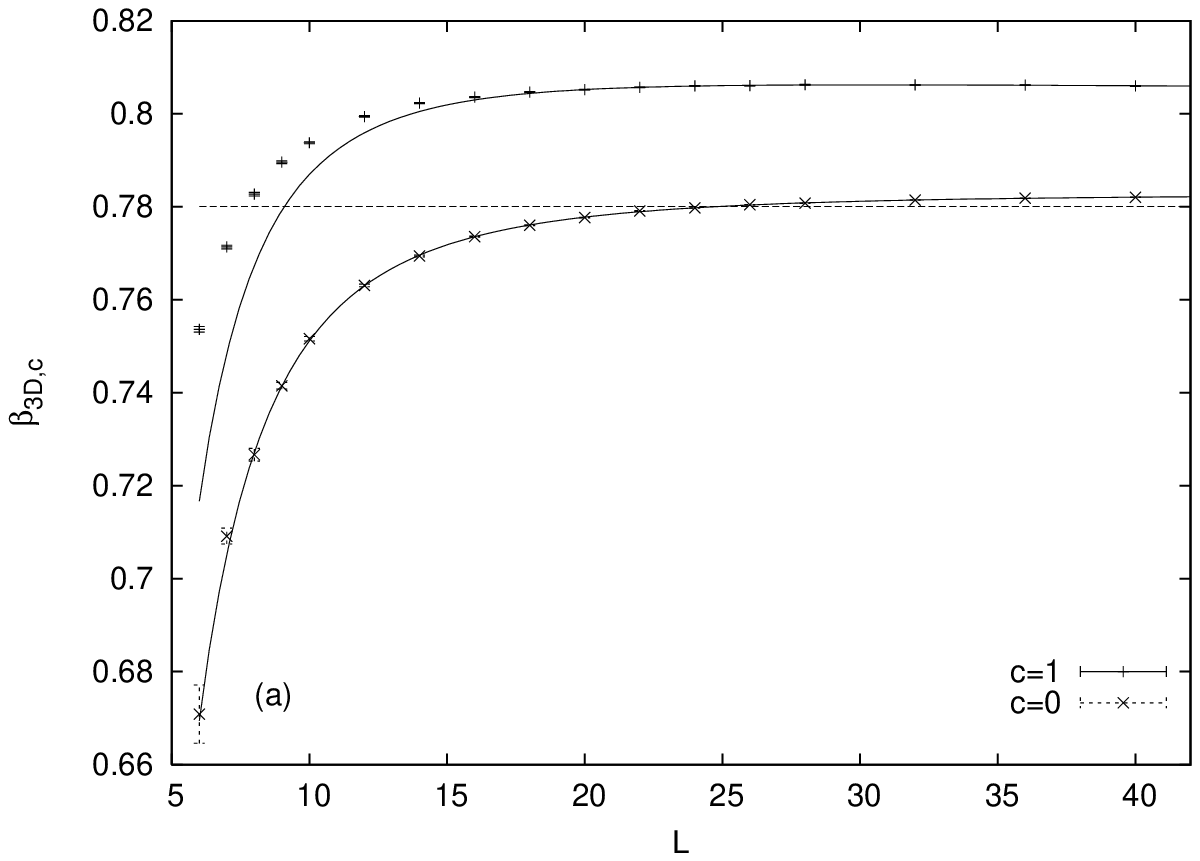,angle=0,height=5.5cm,width=7cm}
\psfig{figure=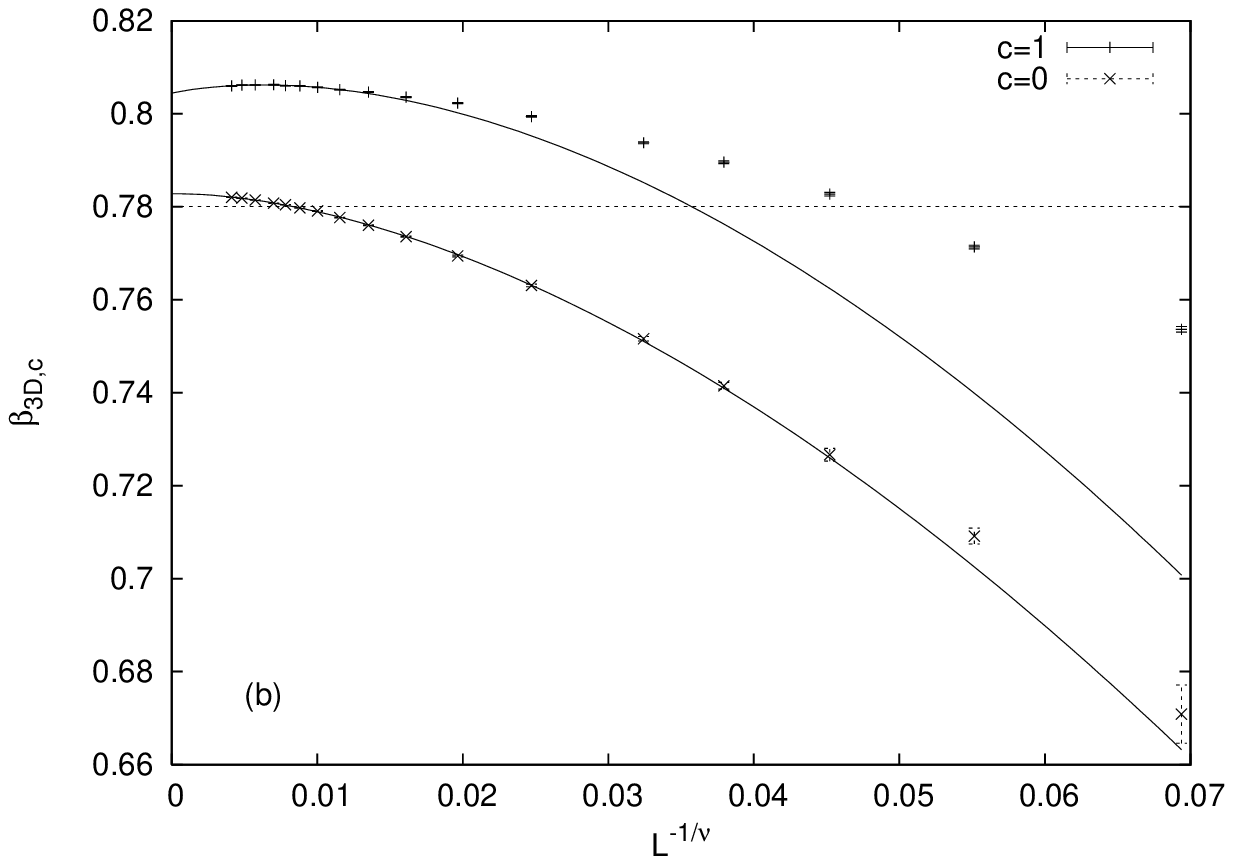,angle=0,height=5.5cm,width=7cm}}
\caption{
\label{fig:p3_1}
Location of the percolation thresholds determined from the 
maximum of susceptibility of ${\cal O}_{\rm 3D}$ for $c=0$ and $c=1$ as a 
function of (a) $L$ and (b) $L^{-1/\nu}$, respectively. The lines indicate fits
according to Eq.~(\ref{fss_beta}) with $\nu$ and $\omega$ fixed according 
to Table~\ref{expos}. The horizontal dashed line shows the thermodynamically
determined critical coupling $\beta_c = 0.780\,08(4)$.
\vspace*{3cm}
}
\end{figure*}

To develop a purely geometric picture of the mechanism governing this transition,
one should thus be more ambitious and also consider the various quantities
$\mathcal{O}_i$ introduced above that focus on the {\em percolative\/} 
properties of the vortex-loop network.
As an example for the various susceptibilities considered, we show in 
Fig.~\ref{fig:p3_0} the susceptibility $\chi_{\rm 3D}$ of ${\cal O}_{\rm 3D}$
for $c=0$ and $c=1$. The resulting scaling behavior of the
maxima locations $\beta_{\rm 3D}(L)$ is depicted in
Fig.~\ref{fig:p3_1}, where the lines indicate fits according to 
Eq.~(\ref{fss_beta}) with exponents fixed again according to Table~\ref{expos}.
We obtain $\beta_{{\rm 3D},c}=0.7824(1)$ with $\chi^2/{\rm dof}=1.14$ 
$(Q=0.32, L\ge 8)$ for $c=0$ and $\beta_{{\rm 3D},c}=0.8042(4)$ with
$\chi^2/{\rm dof}=0.75$ $(Q=0.58, L\ge 20)$ for 
$c=1$. While for the ``stochastic'' rule with $c=0$ the infinite-volume limit 
of  $\beta_{\rm 3D}(L)$ is at least close to $\beta_c$, it is clearly
significantly larger than $\beta_c$ for the fully knotted vortex networks with 
$c=1$.

By repeating the fits for all vortex-network observables and the parameter
$c$ between 0 and 1 in steps of 0.1, we find the results collected in 
Tables~\ref{fss_vol} and \ref{fss_p1}. To check the stability of the fit 
results we performed fits with different lower bounds of the fit range 
$L_{\rm min}$, while the upper bound was always our largest lattice size $L=40$.
For all observables, except for ${\cal O}_{\rm 3D}$, we found a weak dependence
of $\beta_{i,c}$ on the fit range. 
For all five observables we see
that the location of the infinite-volume limit $\beta_{i,c}$ does depend on
the connectivity parameter $c$ used in constructing the vortex loops in 
a statistically significant way. 
With decreasing $c$, the infinite-volume 
extrapolations come closer toward the thermodynamical critical value 
(\ref{eq:thermo_bc}), but even for $c=0$ they clearly do not coincide.

As in Ref.~\onlinecite{kajantie} we found that the percolation points $\beta_{i,c}$
of ${\cal O}_i$ satisfy some inequalities. Because each lattice cube has three 
plaquettes, ${\cal O}_{\rm vol} \ge {\cal O}_{\rm mass}/3$, and it is plausible
that $\langle {\cal O}_{\rm 1D} \rangle \ge  \langle {\cal O}_{\rm 2D} \rangle 
\ge \langle {\cal O}_{\rm 3D} \rangle$. The first relation implies
\begin{equation}
\beta_{{\rm vol},c} \ge \beta_{{\rm mass},c}~.
\end{equation}
Our results collected in Table~\ref{fss_vol} 
are consistent with this inequality.
In addition to this inequality the authors of Ref.~\onlinecite{kajantie} also 
conjectured that $\beta_{{\rm vol},c}= \beta_{{\rm 3D},c}= \beta_{{\rm 1D},c}$.
Our numerical data show that $\beta_{{\rm vol},c}\approx \beta_{{\rm 1D},c}$, but
the other percolation points satisfy only the following inequalities:
\begin{equation}
\beta_{{\rm vol},c}\approx \beta_{{\rm 1D},c} \ge  \beta_{{\rm 2D},c} \ge \beta_{{\rm 3D},c}~,
\end{equation}
cf.\ Table~\ref{fss_p1}.
The reason for this are possibly different corrections to scaling for the different observables.
In the infinite-volume limit all definitions should lead to the same critical point.

These findings are reminiscent of the
percolation behavior of, say, Ising (minority) spin droplets of like spins which
are known to percolate in three dimensions already below the transition
temperature, i.e., $\beta_p > \beta_c$ as for the vortex-loop observables.
Only by breaking bonds between like spins with a certain temperature
dependent probability $p_b^{\rm FK}$ ($= \exp(-2\beta)$), one can tune the thus 
defined Fortuin-Kasteleyn (FK) clusters to percolate at $\beta_c$. With any 
other non-FK probability $0 < p_b < p_b^{\rm FK}$ for 
breaking bonds between like spins it is conceivable that the associated 
percolation point would be located somewhere between $\beta_p$ of the 
geometrical droplets and the thermodynamical (or, equivalently, FK) critical 
point $\beta_c$ (for $p_b > p_b^{\rm FK}$ the percolation transition may even 
vanish altogether). By analogy, our connectivity parameter $c$ seems to play
a similar role for the vortex-loop network as $p_b$ for the spin droplets. 
However, due to the missing analog to the FK representation of the Ising 
model, in the present case of the vortex-loop network, it is not easy to guess
a suitable temperature dependence of the parameter $c$ and we hence eluded to
using a systematic variation of $c$ in small constant increments. The other 
important difference to the case of Ising droplets is of course the long-range
interaction between vortex-line elements which certainly puts the sketched 
analogy to Ising droplets on quite an uncertain and speculative footing.

\begin{figure*}[t]
\centerline{ \psfig{figure=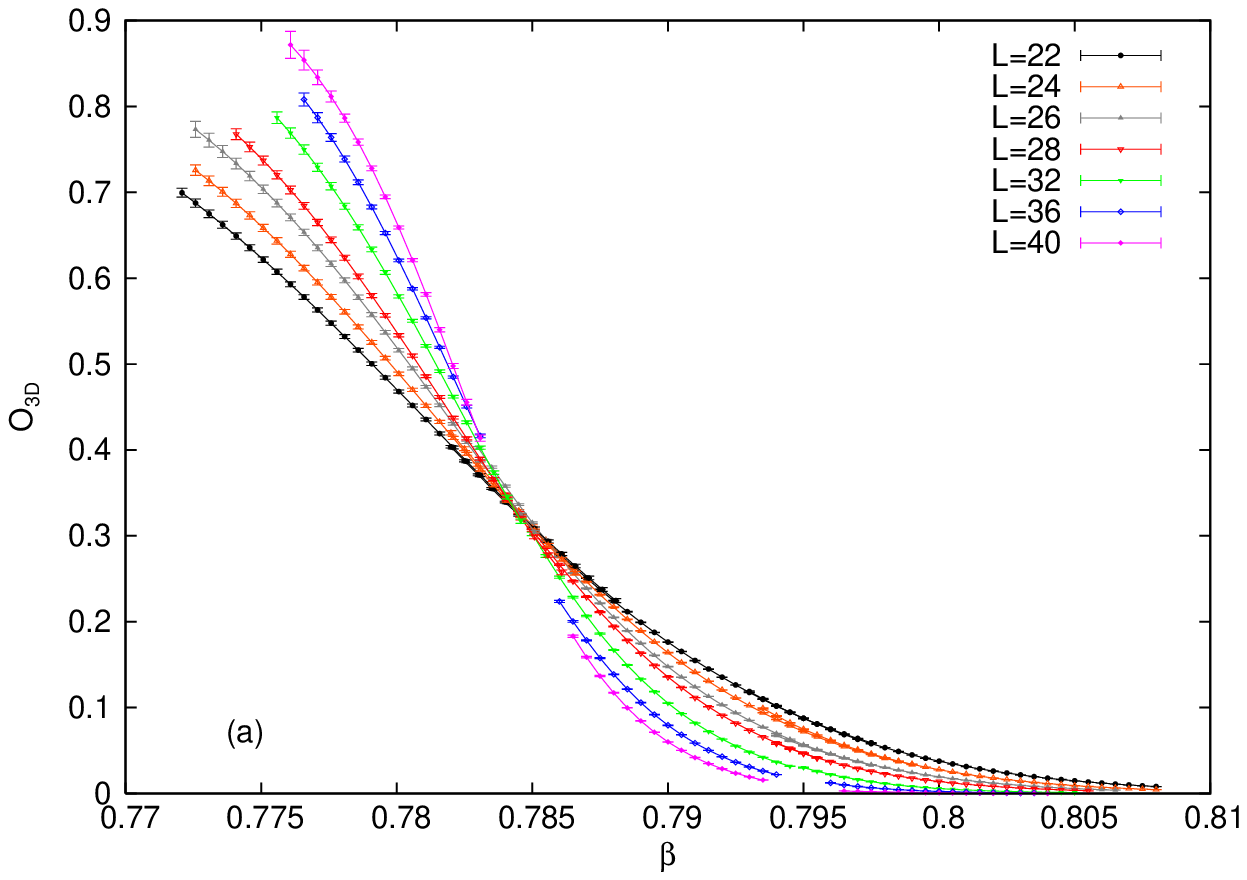,height=5.5cm,width=7cm}
\psfig{figure=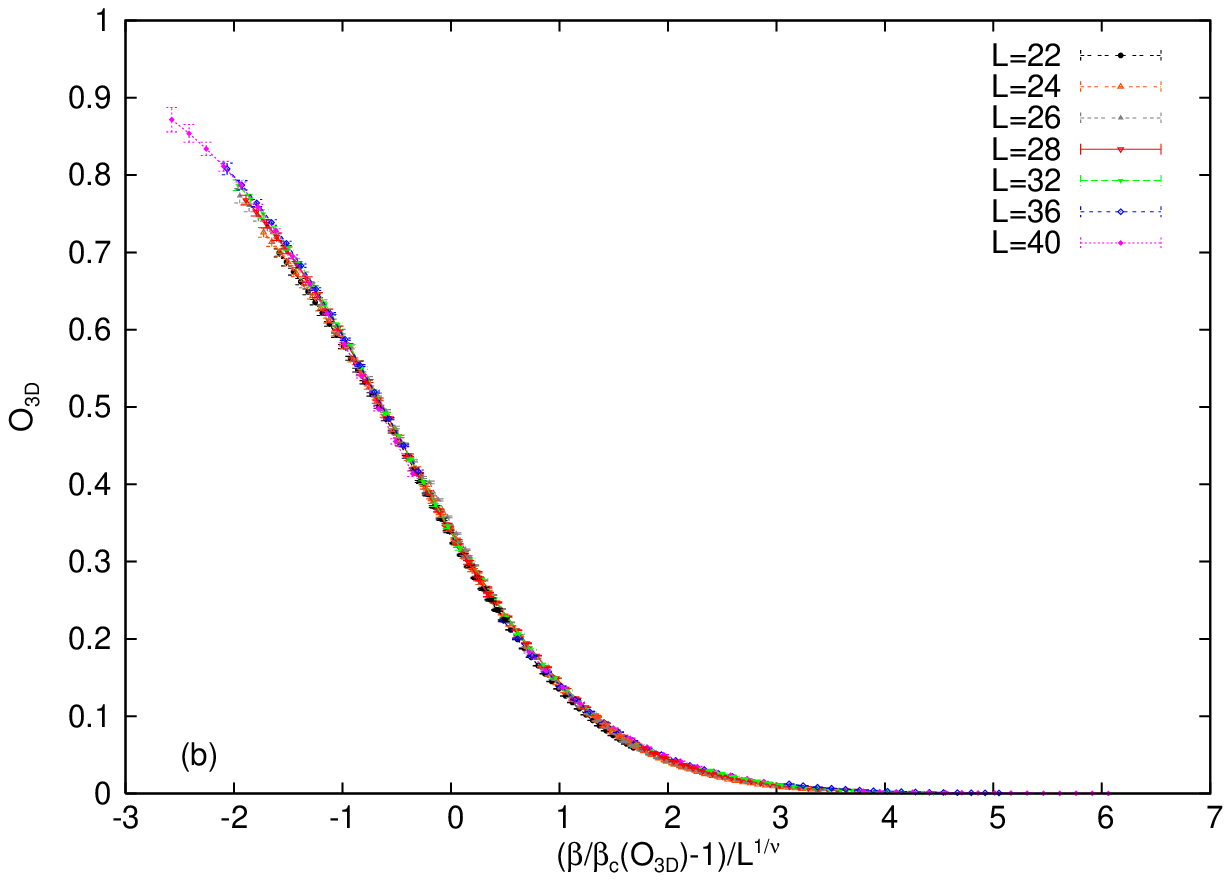,height=5.5cm,width=7cm}}  
\caption{ \label{res_O3}
(Color online) (a) ${\cal O}_{\rm 3D}$ as a function of inverse temperature 
$\beta = 1/T$ for $c=0$. 
(b) Rescaled data with $\nu$ fixed at the 3D $XY$ model value 
(cf.~Table~\ref{expos}) and choosing $\beta_c({\cal O}_{\rm 3D})=0.7842$ to
be the location of the crossing point in (a) for the best data collapse.
}
\end{figure*}

With these remarks in mind we nevertheless performed tests whether at least 
for $c=0$ the critical behavior of the vortex-loop network may consistently
be described by the three-dimensional $XY$ model universality class. 
As an example for a quantity that is {\em a priori\/}
expected to behave as a percolation probability we picked again the
quantity $\mathcal{O}_{\rm 3D}$ for which the susceptibility was already
shown in Fig.~\ref{fig:p3_0}. As is demonstrated in Fig.~\ref{res_O3}(a)
for the case $c=0$, by plotting the
raw data of $\mathcal{O}_{\rm 3D}$ as a function of $\beta$ for the various
lattice sizes, one obtains a clear crossing point so that the interpretation
of $\mathcal{O}_{\rm 3D}$ as percolation probability is 
nicely confirmed. To test the scaling behavior we rescaled
the raw data in the FSS master plot shown in Fig.~\ref{res_O3}(b), where the 
critical exponent $\nu$ has the $XY$ model value given in Table~\ref{expos} and
$\beta_c (\mathcal{O}_{\rm 3D})=0.7842$ was independently determined by 
optimizing the data collapse, i.e., virtually this is the location of the
crossing point in Fig.~\ref{res_O3}(a). The collapse turns out to be quite
sharp which we explicitly judged by comparison with similar plots
for standard bond and site percolation (using there the proper percolation 
exponent, of course). For $c>0$ we found also a sharp data collapse, but for
a monotonically increasing exponent $\nu$, which is for large $c$ values
compatible with the percolation critical exponent $\nu=0.8765(16)$ on a
three-dimensional simple cubic lattice.~\cite{ballesteros}
One should keep in mind, however, that neither 
$\beta_{{\rm 3D},c}$ as extrapolated from the susceptibility
peaks nor the estimate obtained from the crossing point in Fig.~\ref{res_O3}(a) is
compatible with $\beta_c$.

\begin{figure*}[h]
\centerline{ \psfig{figure=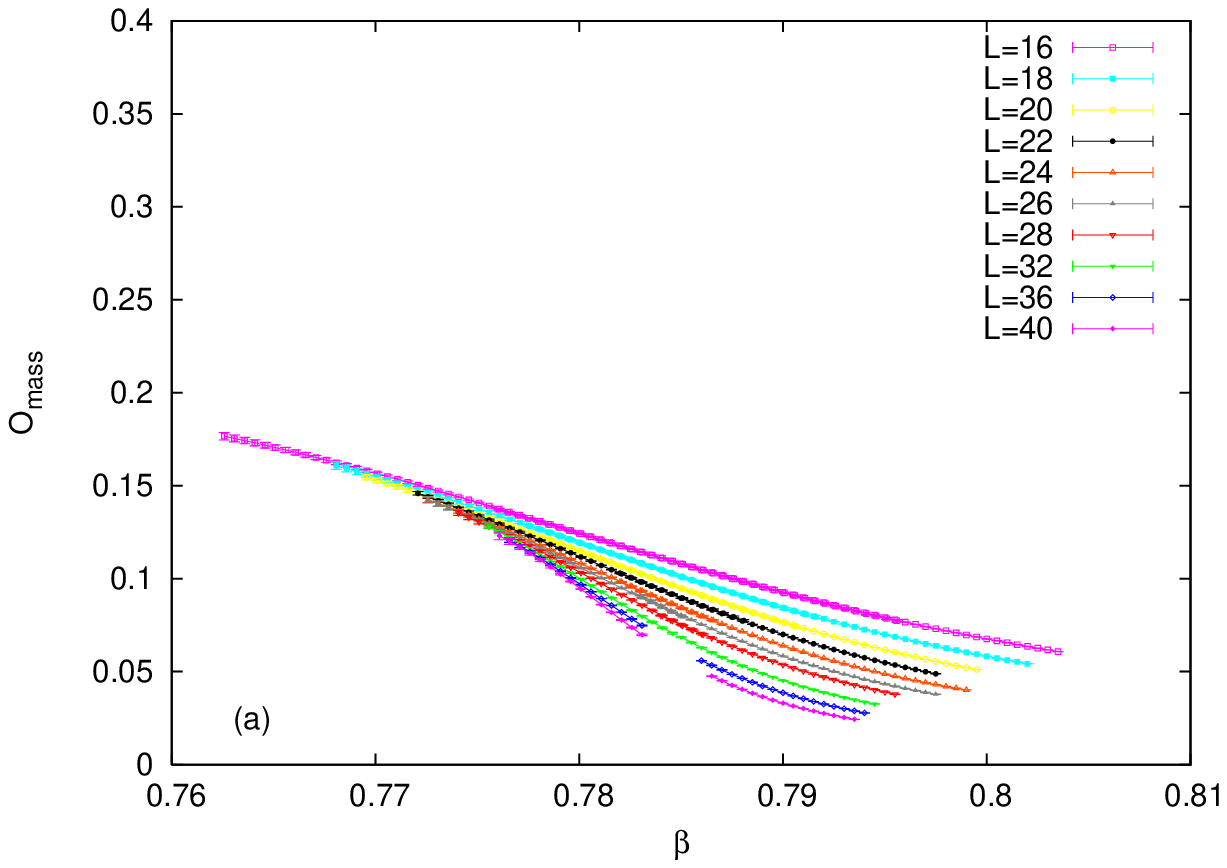,height=5.5cm,width=7cm}
\psfig{figure=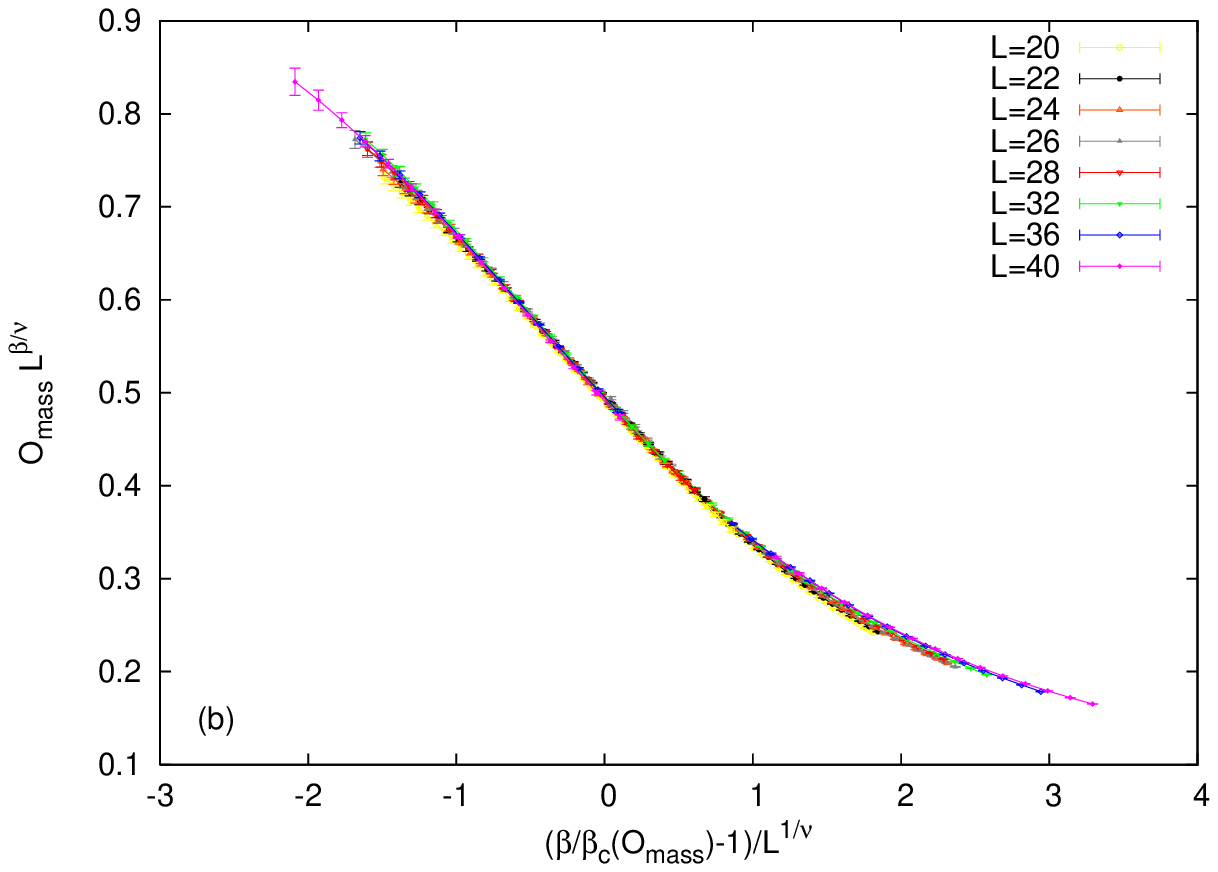,height=5.5cm,width=7cm}}
\caption{ \label{res_0}
(Color online) (a) Mass of vortex loops as a function of inverse temperature 
$\beta = 1/T$ for $c=0$. 
(b) Rescaled data assuming 3D $XY$ model critical exponents (cf.~Table~\ref{expos})
and adjusting $\beta_c({\cal O}_{\rm mass})=0.782\,75$ for the best data collapse.
}
\end{figure*}

Next we looked at ${\cal O}_{\rm mass}$ which {\em a priori\/} is expected to 
behave like a percolation strength, that is similarly to the magnetization 
with an inverted $\beta$ axis. The plot of the raw data for $c=0$ as a
function of $\beta$ in Fig.~\ref{res_0}(a) indeed seems to confirm this 
expectation. To test the scaling properties we show in Fig.~\ref{res_0}(b)
the corresponding FSS master plot, where the critical exponents $\nu$ and 
$\beta$ are again fixed to their $XY$ model values (cf.\ Table~\ref{expos})
and $\beta_c({\cal O}_{\rm mass})=0.782\,75$ was determined by optimizing the
data collapse. Also this collapse is comparatively sharp. Even though the
thus obtained value for $\beta_c({\cal O}_{\rm mass})$ is consistent within 
error bars with the FSS value in Table~\ref{fss_vol} obtained from the 
susceptibility maxima locations (but even further away from $\beta_c$), we 
found a visible spread of the rescaled curves when the latter value was 
used and kept fixed. Similarly, assuming both $XY$ model exponents
{\em and\/} $\beta_c({\cal O}_{\rm mass})=\beta_c$ does {\em not\/} produce
a satisfactory data collapse. Thus for both observables, $\mathcal{O}_{\rm 3D}$ 
and ${\cal O}_{\rm mass}$, we obtain nice FSS scaling plots at $c=0$ compatible 
with $XY$ model critical exponents, but a ``wrong'' critical coupling.

\begin{figure*}[b]
\centerline{ \psfig{figure=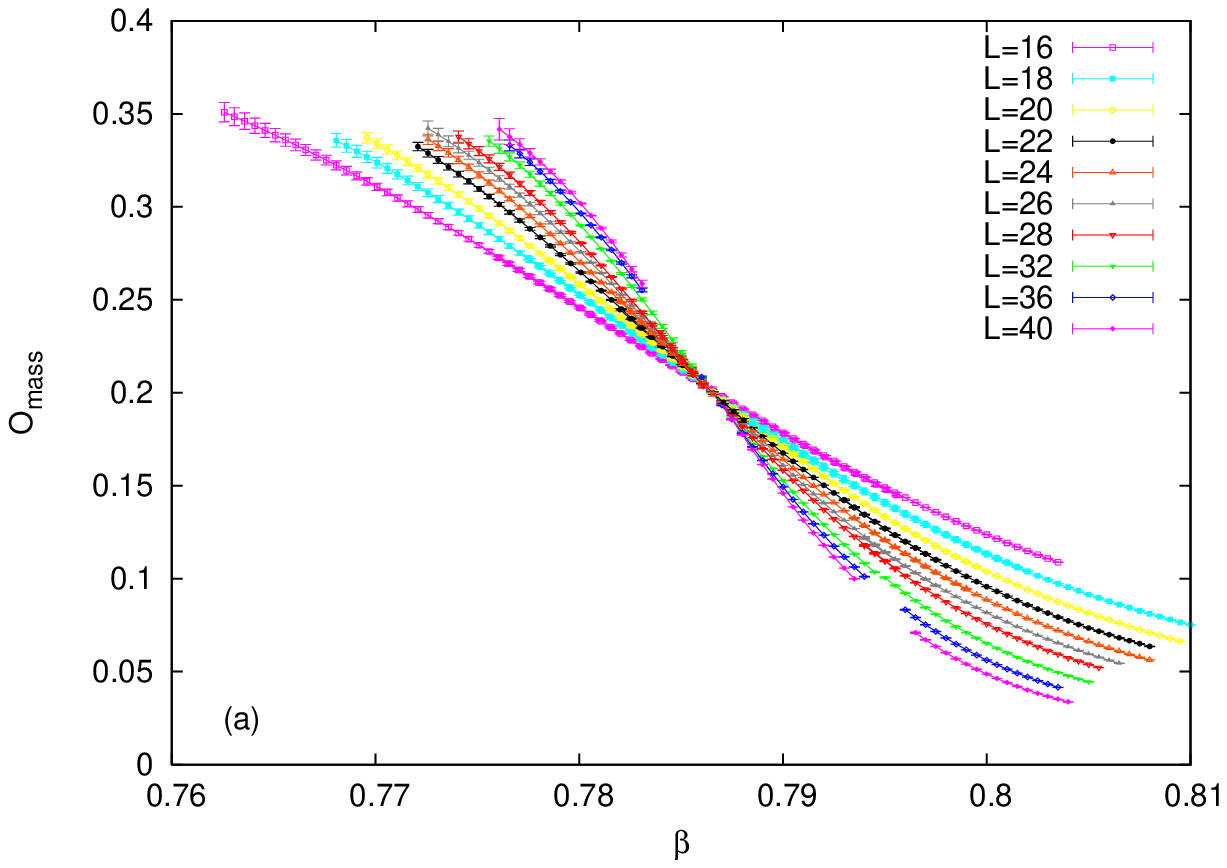,height=5.5cm,width=7cm}
\psfig{figure=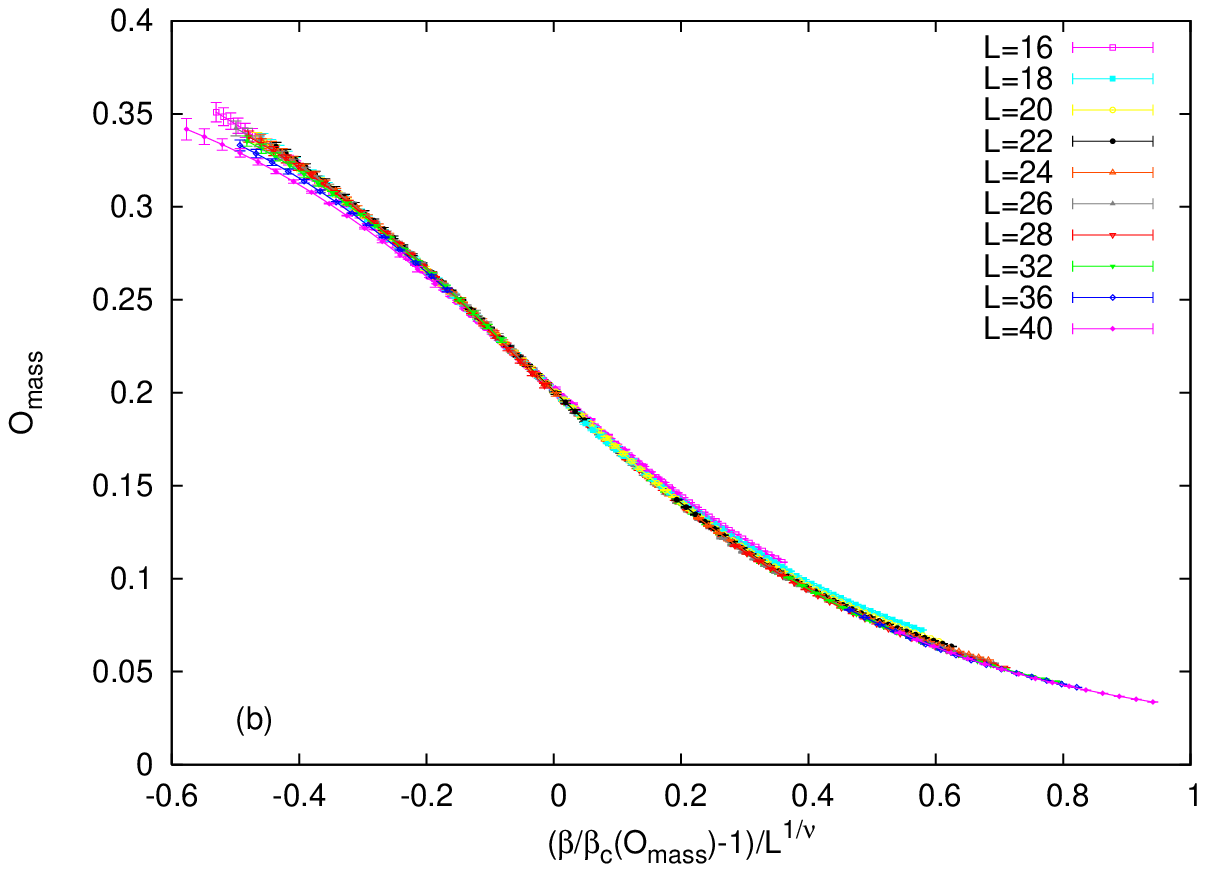,height=5.5cm,width=7cm}}
\caption{ \label{res_0.4}
(Color online) Similar plot as in Fig.~\ref{res_0} for $c=0.4$. Here both
$\beta_c({\cal O}_{\rm mass})=0.786\,46$ and 
$\nu=0.98$ are adjusted to achieve a good data collapse. Note that in 
contrast to  Fig.~\ref{res_0}(b), the $y$ axis in (b) is {\em not\/} rescaled.
}
\end{figure*}

Surprisingly, when using $c\ne0$ for constructing the vortex loops,
${\cal O}_{\rm mass}$ shows a completely different behavior. As example,
we show in Fig.~\ref{res_0.4} our data for the case $c=0.4$. Already by 
looking at the raw data in Fig.~\ref{res_0.4}(a), it is obvious that, 
for $c \ne 0$, the mass of vortex loops no longer behaves as a 
percolation {\em strength\/} (i.e., magnetization); rather it resembles pretty
much the percolation {\em probability\/} ${\cal O}_{\rm3D}$. From the crossing 
point of the curves we get $\beta({\cal O}_{\rm mass})=0.786\,46$.
Using this value we get in the FSS master plot shown in Fig.~\ref{res_0.4}(b) 
a nice data collapse for $\nu=0.98$, but in contrast to the $c=0$ case  now 
only when the $y$ axis is {\em not\/} rescaled. We repeated this analysis for 
all values $c>0$ and found a monotonically increasing exponent $\nu$ 
from $\nu \approx 0.74$ for $c=0.1$ to $\nu \approx 1.82$ for $c=1.0$, which 
appears quite nonsensical. A precise determination of the critical exponents
as a function of $c$ was not our aim here and anyway, due to the rather 
small lattice sizes studied, also not feasible. Still, this strange behavior 
clearly calls for an explanation. 

\begin{figure*}[htb]
\vspace*{3cm}
\centerline{ \psfig{figure=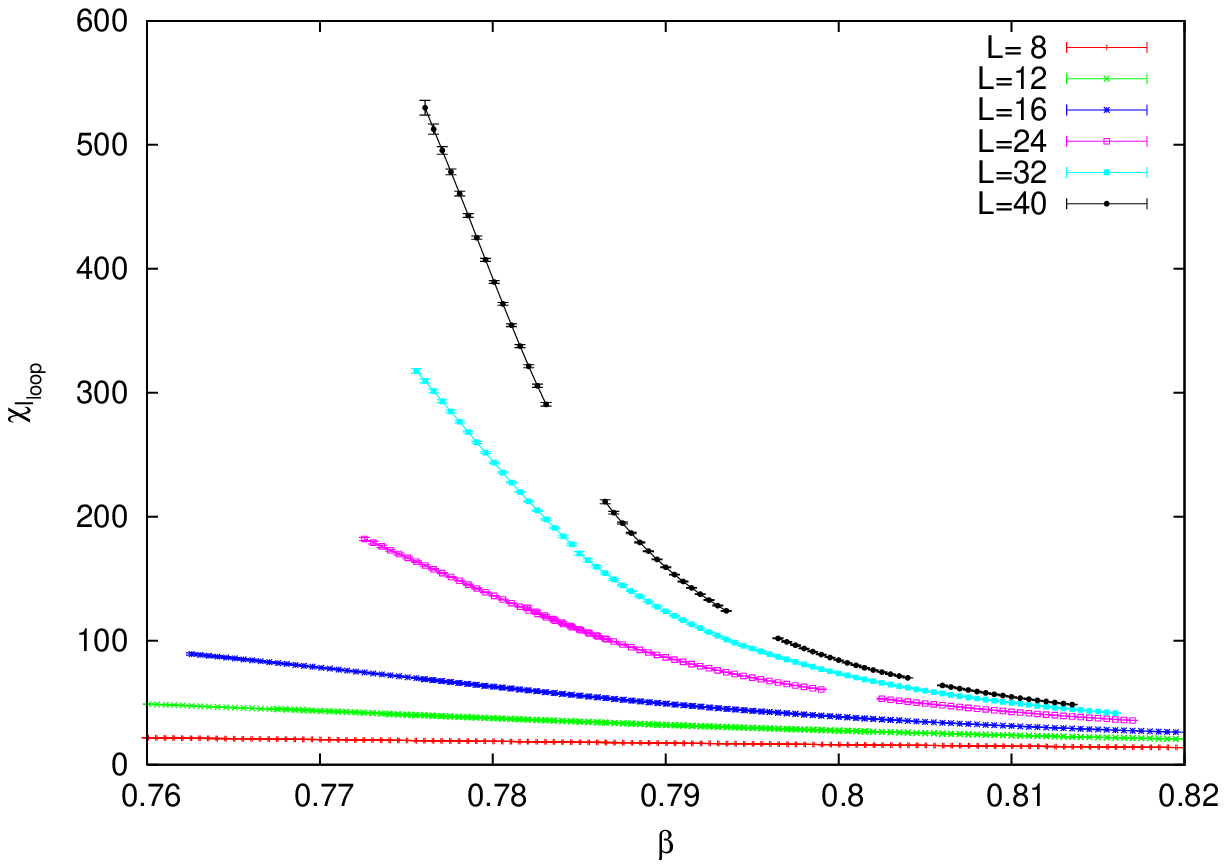,height=5cm,width=6cm}
\psfig{figure=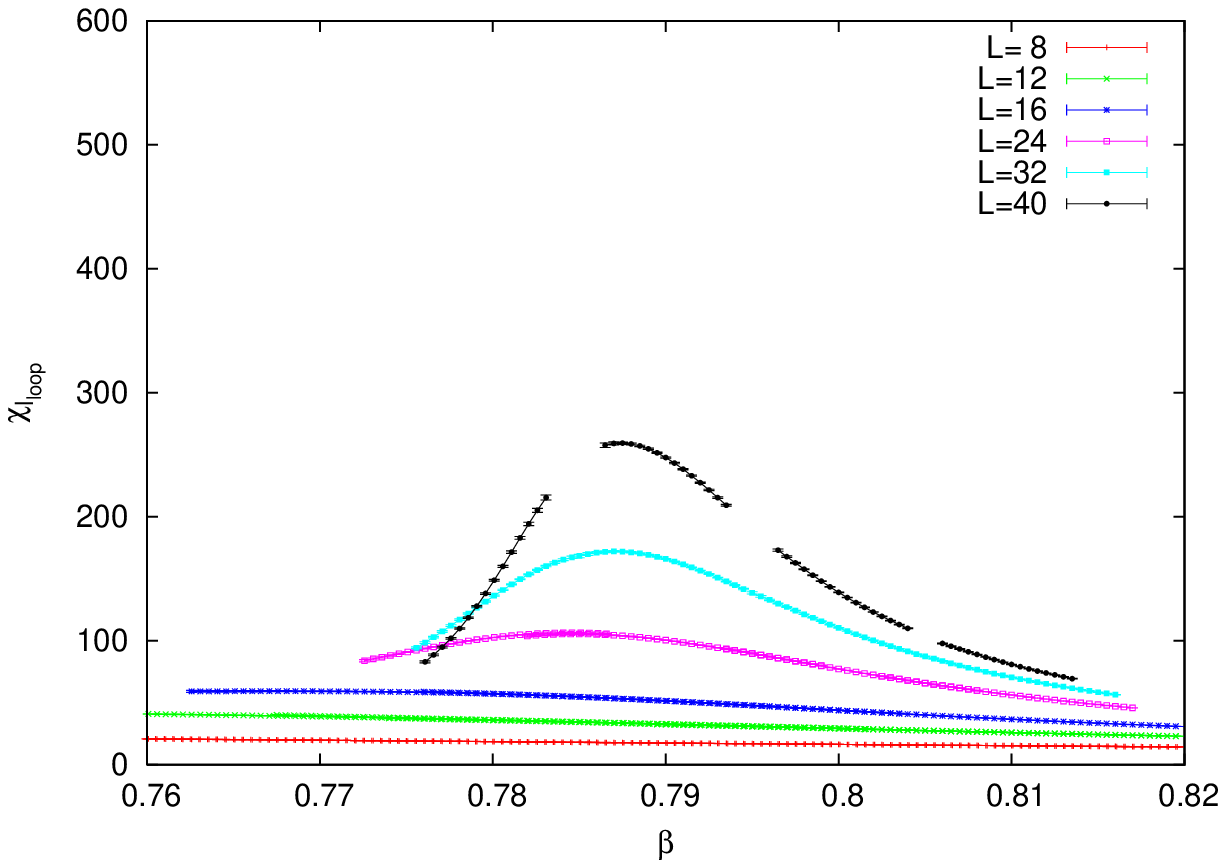,height=5cm,width=6cm}
\psfig{figure=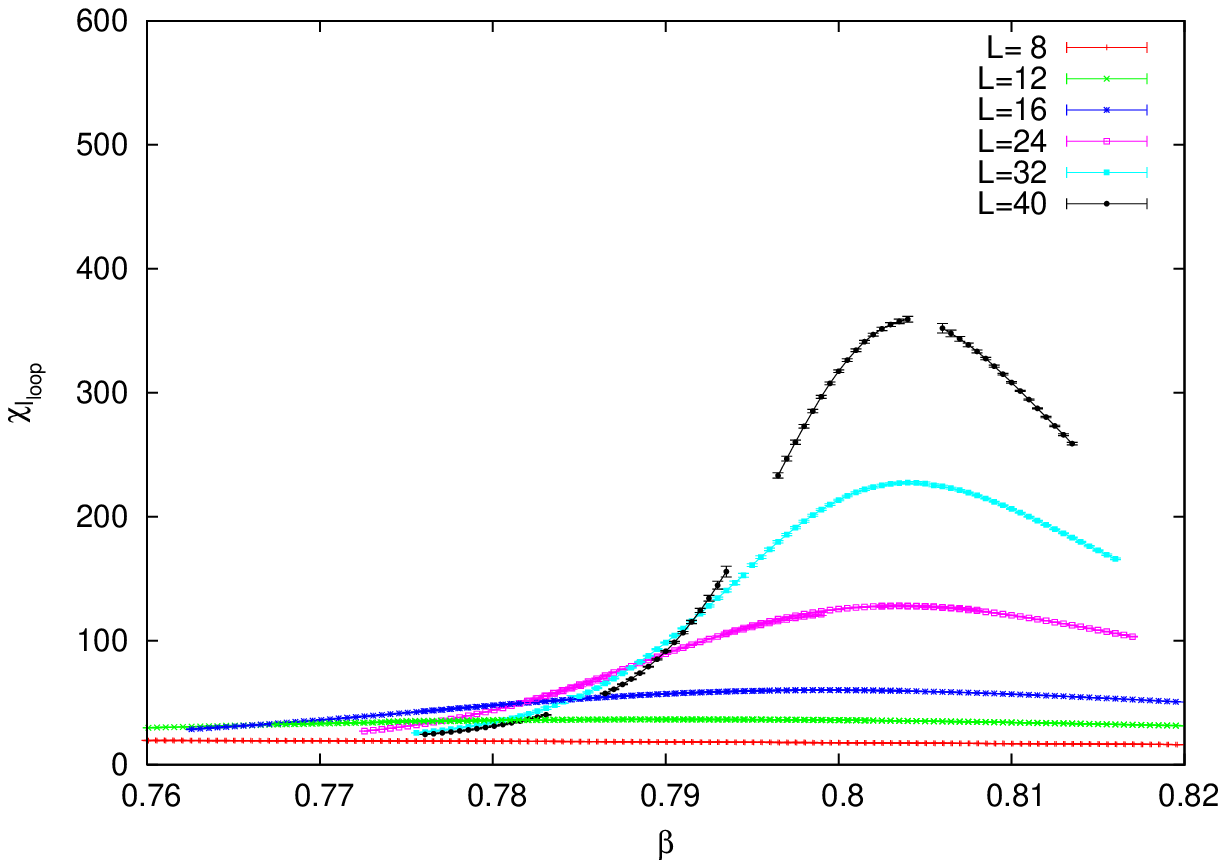,height=5cm,width=6cm}}
\caption{ \label{res_Per}
(Color online) Average loop size $\chi_{l_{\rm loop}}$ as a function of 
inverse temperature $\beta = 1/T$ for $c=0$, $0.2$, and $1$ (from left to right). 
}
\end{figure*}

The variance definition (\ref{eq:variances}) of the susceptibilities studied
so far is quite unusual in percolation theory. We have therefore also
investigated the standard percolation definition for the average loop size
$\chi_{l_{\rm loop}}$ (as seen at a given link of the lattice) which is
expected to scale as the variances defined above. In terms of the loop-length
distribution $P(l_{\rm loop})$ it is given as~\cite{stauffer}
\begin{equation}
\chi_L=\frac{\sum'_{\rm loop} l_{\rm loop}^2 P(l_{\rm loop})}{\sum'_{\rm loop} l_{\rm loop} P(l_{\rm loop})}~, 
\end{equation}
where the prime on the sum is to indicate that we discard in each measurement
the percolating loop according to the criterion $\mathcal{O}_{\rm 3D}$. For
this observable we also found a clear displacement between the maxima for
different values of the connectivity parameter $c$ and the thermodynamic
transition point.
Unfortunately, the reweighting range for $c=0$ was too narrow for this
observable to allow more detailed analyses, see Fig.~\ref{res_Per}.
For $c>0$ the location of the pseudocritical points of the average loop size
behave similar to the susceptibilities as defined in Eq.~(\ref{eq:variances})
and lead to slightly higher $\beta_c$ values than the thermodynamic one.

\begin{figure*}[htb]
\centerline{\psfig{figure=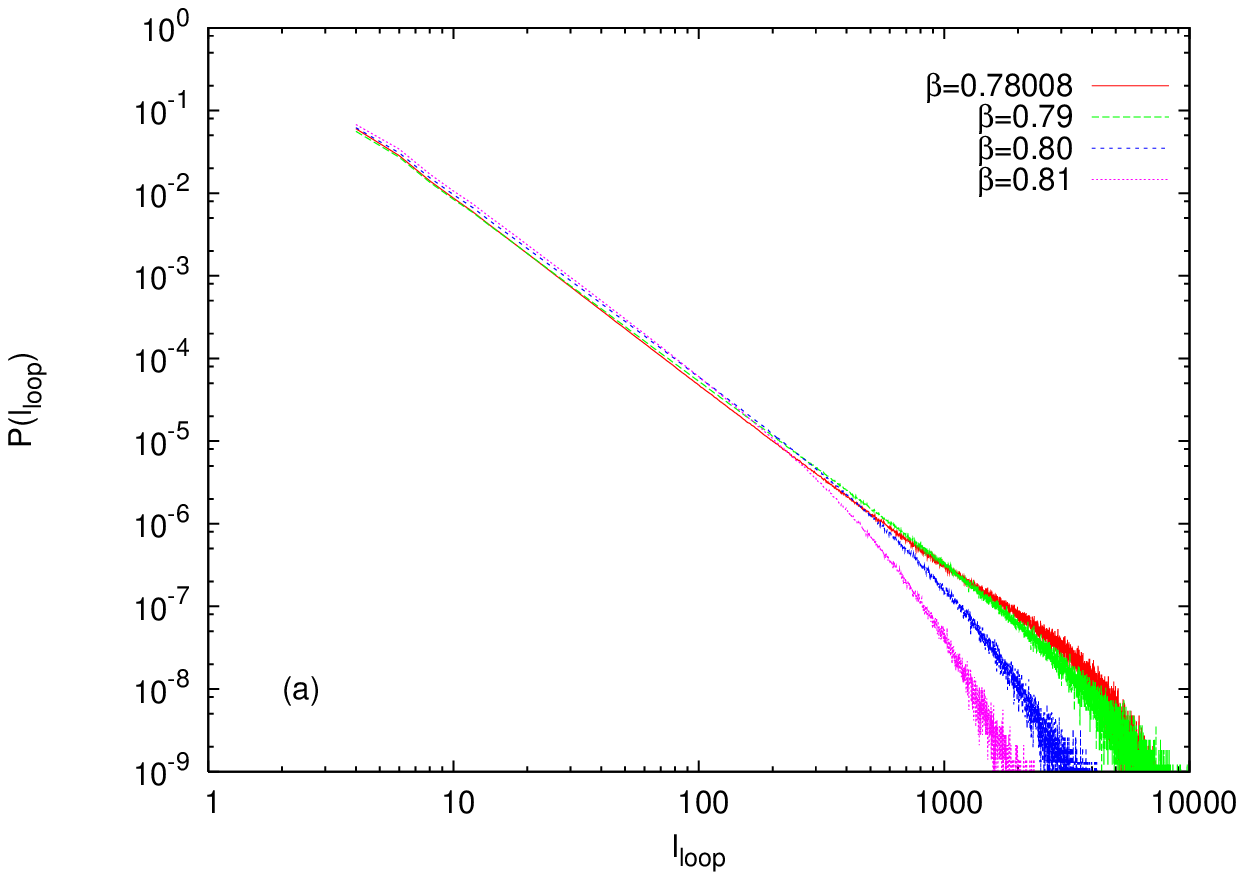,angle=0,height=5.5cm,width=7cm}
\psfig{figure=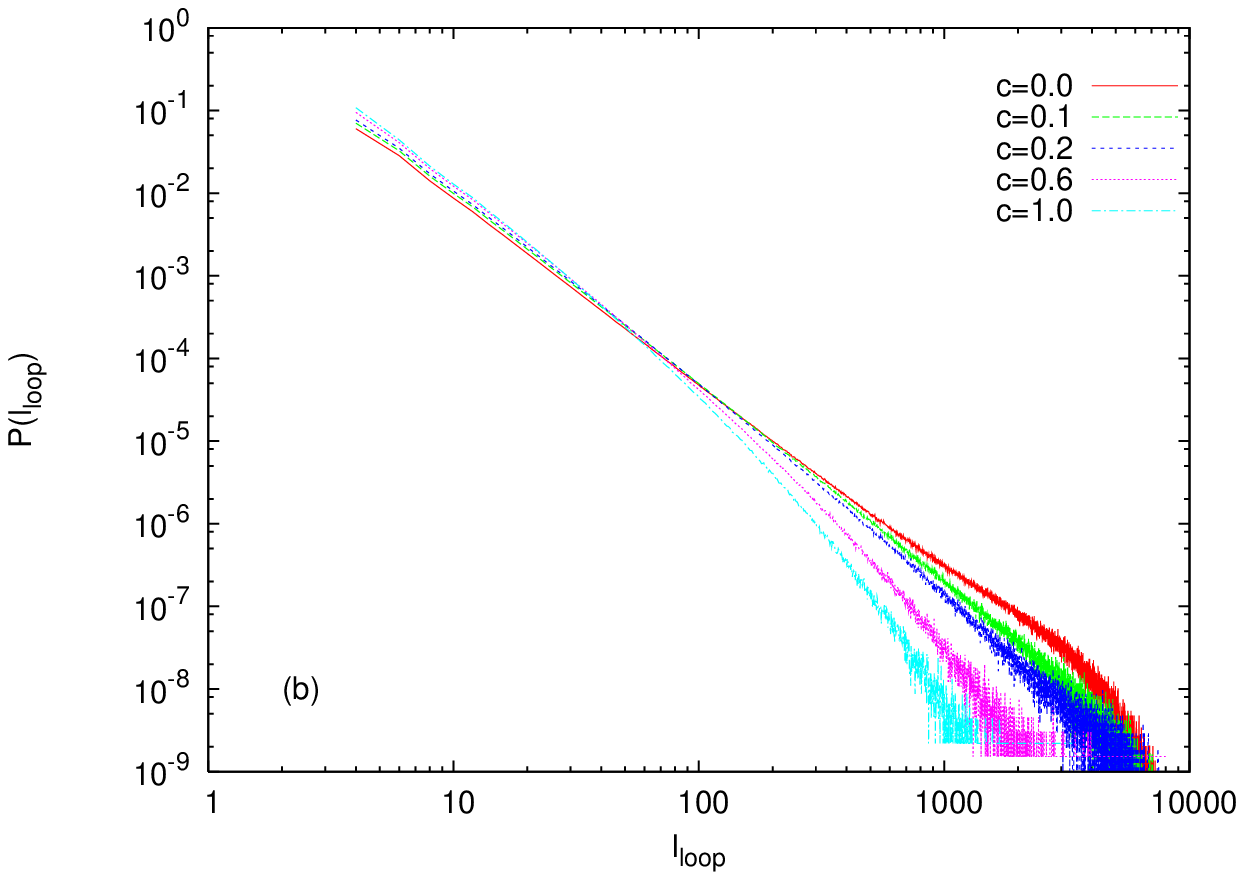,angle=0,height=5.5cm,width=7cm}}
\caption{
\label{fig:histo}
(Color online) Loop-length distribution $P(l_{\rm loop})$ for our largest 
lattice ($L=40$) as a function of the loop length. (a) The ``stochastic'' 
rule  ($c=0$) for various temperatures.
(b) Behavior at the thermodynamic critical point
$\beta=0.780\,08$ for various values of
the connectivity parameter $c$. At $c\approx0.1$ the decay changes from exponential to algebraic implying that
the vortex-line tension $\theta$ vanishes.
}
\vspace*{3cm}
\end{figure*}

Finally, in Fig.~\ref{fig:histo}(a) we show the loop-length distribution 
$P(l_{\rm loop})$ (without the largest loop) as a function of the
loop length $l_{\rm loop}$ for the ``stochastic'' rule ($c=0$) for 
various temperatures and $L=40$. From Eq.~(\ref{eq:P_loop}) one expects 
that the decay changes from
exponential to algebraic at $\beta_c$, because of the vanishing of the
vortex-line tension $\theta$. Also from this analysis we found that the 
percolation transition takes place at a slightly higher $\beta$ value 
than the thermodynamic one. We performed fits according to 
$P(l_{\rm loop})\sim l_{\rm loop}^{-\tau}$ and found the best results 
for $\beta=0.79$ where $\tau \approx 2.2 < 5/2$, see Table~\ref{tau}. 
We want to note that this $\beta$ 
value is for our largest lattice and we only examined the 
loop-length distributions at the $\beta$ values used for the simulation.
To determine the percolation transition and also the line tension with 
the help of the loop-length distributions one would need a finer 
temperature spacing. At $\beta_c$ of the thermodynamic transition 
we looked at the change of the decay of the distributions as a function 
of the connectivity parameter $c$ for $L=40$, see Fig.~\ref{fig:histo}(b). 
Here we found the best agreement with an algebraic decay for $c=0.1$ with 
$\tau=2.348(1)$ and $\chi^2/{\rm dof}=2.9$. 
From this observation and the fact that we found a pronounced peak for the 
largest loops in the distribution at $c=0.1$ and no peak at $c=0.0$, we 
conclude that the line tension vanishes close to $c=0.0$.

\section{Conclusion and Outlook} \label{conclusion}

In this paper we have found for the three-dimensional complex
Ginzburg-Landau field theory that the geometrically defined percolation 
transition of the vortex-loop network is close to the thermodynamic phase 
transition point, but does {\em not\/} coincide with it for any connectivity 
definition we have studied. 
Our results for the connectivity parameter $c \in [0,1]$ extend the claim
of Ref.~\onlinecite{kajantie} for the three-dimensional $XY$ spin model that 
neither the ``maximal'' ($c=1$) nor the
``stochastic'' rule ($c=0$) used for constructing macroscopic vortex
loops does reflect the properties of the true phase transition in a strict 
sense.~\cite{sudbo1}
Nevertheless it may be possible to bring the 
percolation transition closer to the thermodynamic one by using different 
vortex-loop network definitions, e.g., using a temperature-dependent or
a size-dependent connectivity parameter in analogy to the Fortuin-Kasteleyn
definition for spin clusters. To verify this presumption 
would be an interesting future project, but thereby one should first 
investigate the $XY$ model which is much less CPU time-consuming. 

\section{Acknowledgments}

We would like to thank Adriaan Schakel for many useful discussions.
Disorder: from membranes to quantum gravity'' for a postdoctoral grant. 
Financial support by the Deutsche Forschungsgemeinschaft (DFG) under grant No.\ JA 483/17-3 and
the German-Israel-Foundation (GIF) under contract No.\ I-653-181.14/1999 is also gratefully
acknowledged.

\begin{table*}[h]
\caption{\label{fss_vol}FSS fits according to Eq.~(\ref{fss_beta}) in the range 
$L_{\rm min}$ to $L_{\rm max}=40$ for the mass and volume order parameter 
of the vortex loops for various values of $c$ ($c=1$: maximally knotted),
with exponents $\nu$ and $\omega$ fixed according to Table~\ref{expos}.
The thermodynamic transition is at $\beta_c=0.780\,08(4)$ 
(Ref.~\onlinecite{ebwj_PRB}).}
  \begin{center}
     \begin{tabular}[b]{clccccccc}\hline \hline
      \multicolumn{1}{c}{}&\multicolumn{4}{c}{${\cal O}_{\rm mass}$}&\multicolumn{4}{c}{${\cal O}_{\rm vol}$}\\ \hline
         \makebox[1cm][c]{$c$} & \makebox[1.3cm][c]{$\beta_c$}& \makebox[1cm][c]{$L_{\rm min}$}
          &\makebox[1cm][c]{$\chi^2/{\rm dof}$}&\makebox[1cm][c]{$Q$}
          &\makebox[1.3cm][c]{$\beta_c$}& \makebox[1cm][c]{$L_{\rm min}$}
          &\makebox[1cm][c]{$\chi^2/{\rm dof}$}&\makebox[1cm][c]{$Q$}\\ \hline
          1.0 & 0.8017(8) &20&1.53&0.17 &0.8072(4)&12&1.86&0.05\\
          0.9 & 0.8016(6) &16&1.66&0.11 &0.8048(3)&14&1.08&0.37\\
          0.8 & 0.7996(9) &16&1.55&0.14 &0.8028(3)&14&1.40&0.19\\
          0.7 & 0.7976(5) &14&1.34&0.21 &0.8005(4)&16&1.14&0.33\\
          0.6 & 0.7970(5) &16&1.10&0.36 &0.7995(5)&14&0.57&0.80\\
          0.5 & 0.7954(6) &18&1.32&0.24 &0.7963(3)&16&1.23&0.28\\
          0.4 & 0.7912(5) &14&0.94&0.48 &0.7938(4)&18&0.82&0.55\\
          0.3 & 0.7887(9) &16&1.32&0.23 &0.7924(4)&18&1.79&0.09\\
          0.2 & 0.7856(3) &~6&1.62&0.07 &0.7907(4)&12&0.28&0.98\\
          0.1 & 0.7834(4) &10&0.37&0.96 &0.7875(2)&12&0.94&0.49\\
          0.0 & 0.7811(39)&10&1.26&0.24 &0.7834(3)&16&0.92&0.48\\
	  \hline \hline
      \end{tabular}
   \end{center}
\end{table*}

\begin{table*}[htb]
\caption{\label{fss_p1} Same as in Table~\ref{fss_vol} for the vortex parameters ${\cal O}_{\rm 1D}$,
 ${\cal O}_{\rm 2D}$, and ${\cal O}_{\rm 3D}$.}
  \begin{center}
     \begin{tabular}{ccccccccclccc}\hline\hline
      \multicolumn{1}{c}{}&\multicolumn{4}{c}{${\cal O}_{\rm 1D}$}&\multicolumn{4}{c}{${\cal O}_{\rm 2D}$}
       &\multicolumn{4}{c}{${\cal O}_{\rm 3D}$}\\ \hline
         \makebox[1cm][c]{$c$} & \makebox[1.3cm][c]{$\beta_c$}& \makebox[1cm][c]{$L_{\rm min}$}
          &\makebox[1cm][c]{$\chi^2/{\rm dof}$}&\makebox[1cm][c]{$Q$}
          &\makebox[1.3cm][c]{$\beta_c$}& \makebox[1cm][c]{$L_{\rm min}$}
          &\makebox[1cm][c]{$\chi^2/{\rm dof}$}&\makebox[1cm][c]{$Q$}
          &\makebox[1.3cm][c]{$\beta_c$}& \makebox[1cm][c]{$L_{\rm min}$}
          &\makebox[1cm][c]{$\chi^2/{\rm dof}$}&\makebox[1cm][c]{$Q$}\\ \hline
          1.0&0.8066(5)&16&0.68&0.68&0.8051(3)&20&0.86&0.50&0.8042(4)&20&0.75&0.58\\
          0.9&0.8048(3)&16&0.68&0.69&0.8032(3)&20&0.72&0.61&0.8027(3)&18&0.37&0.89\\
          0.8&0.8030(2)&16&1.04&0.40&0.8027(4)&14&0.74&0.66&0.8009(2)&18&0.54&0.77\\
          0.7&0.8011(2)&16&0.90&0.50&0.7999(3)&18&1.29&0.26&0.7988(2)&18&0.61&0.72\\
          0.6&0.7992(3)&16&1.67&0.11&0.7976(2)&18&0.44&0.85&0.7968(2)&18&1.20&0.30\\
          0.5&0.7966(3)&16&0.94&0.47&0.7953(2)&18&0.42&0.86&0.7953(2)&18&0.72&0.67\\
          0.4&0.7938(3)&20&1.18&0.32&0.7940(2)&14&1.10&0.36&0.7928(2)&12&0.35&0.96\\
          0.3&0.7919(2)&18&1.11&0.35&0.7915(3)&16&0.66&0.71&0.79037(5)&~6&0.54&0.91\\
          0.2&0.7907(2)&12&1.10&0.37&0.7890(2)&16&0.68&0.69&0.7881(1)&~6&0.92&0.53\\
          0.1&0.7868(3)&18&1.38&0.21&0.7861(2)&18&0.49&0.81&0.7857(2)&~6&0.60&0.86\\
          0.0&0.7837(2)&12&1.84&0.08&0.7824(5)&18&0.28&0.94&0.7824(1)&~8&1.14&0.32\\ \hline\hline
      \end{tabular}
   \end{center}
\end{table*}

\begin{table*}[htb]
  \caption{\label{tau} Results for the Fisher exponent $\tau$ 
for the ``stochastic'' definition ($c=0$) at three inverse temperatures
$\beta = 1/T$,
assuming an algebraic decay of the loop-length distribution close to criticality. 
   }
 \begin{center}
  \begin{tabular}{cccc}\hline\hline
   \makebox[1cm][c]{$\beta$} & \makebox[1.5cm][c]{fit range}& \makebox[1cm][c]{$\tau$}& \makebox[1cm][c]{$\chi^2/{\rm dof}$}\\ \hline
    0.78008&$20 - 500$&2.261(1)&3.1\\
    0.79&$20 - 800$&2.201(1)&1.4\\
    0.80&$20 - 500$&2.275(1)&26\\ \hline\hline
  \end{tabular}
 \end{center}
\end{table*}


\end{document}